\documentclass{llncs}
\usepackage[T1]{fontenc}

\usepackage{verbatim} 
\usepackage{alltt}
\usepackage{url}
\usepackage{amsmath}
\usepackage{ifthen}
\usepackage{amssymb}
\usepackage{fancyhdr}
\usepackage{framed} 
\usepackage{color}
\usepackage{stmaryrd}
\usepackage{multirow}
\usepackage{latexsym}
\usepackage{tikz}

\usetikzlibrary{arrows}
\tikzstyle{oval}=[circle,draw,inner sep=0pt,minimum size=10mm]
\tikzstyle{rond}=[circle,draw,inner sep=0pt,minimum size=2mm]

\usepackage{graphicx}

\definecolor{flash}{rgb}{0.80,0,0}
\definecolor{flashb}{rgb}{0,0,0.80}

\newcommand{\coqbinfd}{{\sf CoqbinFD}}
\newcommand{\alldiff}{{\sf AllDifferent}}

\newenvironment{compact-enumerate}{
\begin{enumerate}
     \setlength{\itemsep}{1pt}
     \setlength{\parskip}{0pt}
     \setlength{\parsep}{0pt}}
{\end{enumerate}
}

\newenvironment{compact-itemize}{
\begin{itemize}
     \setlength{\itemsep}{1pt}
     \setlength{\parskip}{0pt}
     \setlength{\parsep}{0pt}}
{\end{itemize}
}

\def\_{\kern.08em\vbox{\hrule width.35em height.6pt}\kern.08em}

\newcommand{\coqdockw}[1]{\texttt{#1}}

\newcommand{\coqdocvar}[1]{\texttt{#1}}

\newenvironment{coqdoccode}{}{}

\newlength{\coqdocbaseindent}
\newcommand{\coqdocnoindent}{\noindent\kern\coqdocbaseindent}
\newcommand{\coqdocindent}[1]{\noindent\kern\coqdocbaseindent\noindent\kern#1}
\newcommand{\coqdoceol}{\hspace*{\fill}\setlength\parskip{0pt}\par}
\newcommand{\coqdocemptyline}{\vskip 0.4em plus 0.1em minus 0.1em}

\definecolor{varpurple}{rgb}{0.4,0,0.4}
\definecolor{constrmaroon}{rgb}{0.6,0,0}
\definecolor{defgreen}{rgb}{0,0.4,0}
\definecolor{indblue}{rgb}{0,0,0.8}
\definecolor{kwred}{rgb}{0.8,0.1,0.1}

\def\coqdocvarcolor{varpurple}
\def\coqdockwcolor{kwred}

\def\coqdockw#1{{\color{\coqdockwcolor}{\texttt{#1}}}}
\def\coqdocvar#1{{\color{\coqdocvarcolor}{\textit{#1}}}}

\begin{document}

\title{Formally Verified Transformation of Non-binary Constraints into Binary Constraints 
}
\author{Catherine Dubois}
\institute{
Samovar, ENSIIE, \'Evry, France \\
 \email{catherine.dubois@ensiie.fr} \\
 }

\authorrunning{C. Dubois}

\titlerunning{
Formally Verified Transformation of Non-binary Constraints into Binary Constraints
}

\maketitle

\def\domain{\texttt{dom}}
\def\infer{\vdash}
\def\Infer{\models}
\def\give{\leadsto}
\def\givebis{\rhd}
 \def\valn{\coqdocvar{value\_n}}
 \def\varn{\coqdocvar{variable\_n}} 
\def\constn{\coqdocvar{constraint\_n}} 
\def\nary{\coqdocvar{Nary}} 
\def\wfn{\coqdocvar{network\_inv\_n}}

\begin{abstract}

It is well known in the Constraint Programming community that any non-binary constraint satisfaction problem  (with finite domains) can be transformed into an equivalent binary one. One of the most well-known translations is the Hidden Variable Encoding. In this paper we formalize this encoding in the proof assistant Coq and prove that any solution  of the binary constraint satisfaction problem  makes it possible to build a solution of the  original problem and vice-versa.  This formal development is used to complete the formally verified constraint solver developed in Coq by 
 Carlier, Dubois and Gotlieb  in 2012, making it a tool able to solve any n-ary constraint satisfaction problem,  The key  of success of the connection between the translator  and the Coq binary solver is the genericity of the latter. 
\end{abstract}

\section{Introduction}
Constraint Programming or  Constraint Satisfaction Problems \cite{handbook} have many real-life applications such as decision making, resource allocation, scheduling, vehicule routing, configuration, planning, program verification, etc.  Models are made of variables, domains which define the possible values of the variables and constraints which restrict the space of solutions. For example, modelling a Sudoku game  requires 9*9 variables representing the different cells, their domain is the interval 1..9 and the constraints  impose that the numbers in the cells must be all different in each column and each line, and that in each square we must find all the numbers from 1 to 9.  Here constraints can be expressed using the specialized n-ary constraint \alldiff{} \cite{Regin94}. 
Complex problems are usually naturally modelled with constraints involving a large number of variables.  Historically, research in this area has focussed on binary constraints, i.e constraints using only two distincts variables. Then some transformations allowing to translate a non-binary problem containing constraints involving more than two variables, into an equivalent binary problem have been proposed, one of them is the Hidden Variable Encoding (HVE) \cite{RossiHVE}, well-known in the Constraint Programming community. In this paper, we formalize this encoding in Coq and prove that it does provide an equivalent encoding, in the sense that any solution of the encoding binary problem can be translated into a solution of the original non-binary problem and vice-versa. Furthermore if the original problem is unsatisfiable, then the encoding is also unsatisfiable and vice-versa. 

This formal development related to HVE is used to extend the formally verified constraint binary solver developed in Coq by 
 Carlier, Dubois and Gotlieb in 2012~\cite{CDG12}, called \coqbinfd, making it a solver able to solve any n-ary  problem. As far as we know, we provide here the first non-binary constraint solver (for finite domains) formally verified, extracted from a Coq development. It can serve as a reference solver for testing other constraint solvers. It can be compared to the  verified LTL model checker developed in Isabelle/HOL  proposed as  a reference implementation in~\cite{EsparzaLNNSS13}. 
It is also a brick of a formal library dedicated to formalize results and classical algorithms about constraints, in the spirit of the project IsaFoL (Isabelle formalisation of Logic)\footnote{\url{https://bitbucket.org/isafol/isafol/wiki/Home}}  which includes e.g.  the formalisation in Isabelle/HOL of a CDCL-based
SAT solver using  efficient imperative data structures~\cite{FleuryBL18}.

In \cite{jfla19}, we have presented such an encoding verified in Coq for ternary constraints only. This intermediate step was helpful to achieve the  n-ary generalization. The two Coq formalisations are close and follow the same process. The main lemmas and theorems are if not identical, very close to each other. The reason why we have first done the ternary case is historical: the translation was implemented in OCaml to encode non-binary arithmetic constraints as a set of ternary constraints (it can always be done as long as only binary and unary operators occur in the non-binary constraint) in order to use  \coqbinfd. Then we decided to push this transformation into Coq and to verify it for finally achieve the formalisation we present in tis paper. The ternary version does not take into account extensional constraints. 
 
The paper is organized as follows. Section \ref{sec:bg} briefly presents the notion of constraint satisfaction problem, the main ingredients of a constraint solver and the Hidden Variable Encoding. Section \ref{sec:hve} describes the Coq formalisation of the Hidden Variable Encoding and highlights the proven properties. Then Section \ref{sec:coqbinfd}  introduces the main characteristics of \coqbinfd. Section \ref{sec:extension} presents the extended solver, obtained by reusing \coqbinfd{} and also some experimentations. 
We conclude in the last section.

\section{Background}\label{sec:bg}
A {\it Constraint Satisfaction Problem} (csp for short) or network of
constraints \cite{AI77} is a triple $(X,D,C)$ where $X$ is a set of
variables, $C$ is a set of constraints over $X$
and $D$ is a  function that associates a finite domain $D(x)$
to each variable $x$ in $X$. 
Constraints are relationships between variables, each
taking a value in their respective domain: constraints restrict possible
values that variables can take.
As often in  CP literature, we assume that constraints are
normalized, meaning that two
distinct constraints cannot hold over exactly the same variables. The
arity of a constraint is the number of its variables (assumed as distinct).
A $n$-ary csp contains $k$-ary constraints with $k \le n$. A csp is said non-binary as soon as it contains a constraint whose arity is strictly greater than 2. We do not consider unary constraints since the constraint can be directly taken into account in the domain. 
A solution is defined  as a total assignment of the csp variables
which satisfies all the constraints simultaneously.

Let us consider as an example the following  non-binary csp $(X, D,
C)$  where 
$X = \{ x_1, x_2, x_3, x_4, x_5, x_6\}$, 
$D(v) = \{0, 1\}$ for all $v$ in $X$ and 
$C = \{c_1 : x_1 + x_2 + x_6 = 1, c_2 : x_1 + x2 - x_3 + x_4 = 1, c_3 : x_4 +
x_5 - x_6 \geq 1, c_4 : x_2 + x_5 - x_6 = 0, c_5 : x_1 \ge x_6  \}$
inspired from~\cite{StergiouW99}.
It has a unique solution   defined as $\{
x_1 \mapsto 1 , x_2 \mapsto 0,  x_3 \mapsto 1, x_4 \mapsto 1, x_5
\mapsto 0, x_6 \mapsto 0\}$. 

A constraint solver  usually alternates propagation and exploration. Propagation prunes the domains of the variables, removing inconsistent values, using the constraints. This step can be decomposed in 2 interleaved routines: filtering that removes inconsistent values from the domains of the variables of  one constraint and propagation that determines the constraints that have to be visited after a filtering step until a fixpoint is reached. Exploration  enumerates values for some variables and may backtrack on these choices if necessary. The propagation step enforces a local consistency property that characterizes some necessary conditions on values to belong to solutions. There exist many different local consistencies, e.g.  arc consistency, path consistency or bound consistency \cite{Bes06}. One of the oldest is Arc Consistency (when applied to binary constraints)  or Generalized Arc Consistency (as a generalization of AC to n-ary constraints). 
Let $c$ be a constraint  of a csp $(X,D, C)$ whose variables are $x_1$, $x_2$ \ldots  $x_k$. The constraint $c$ is (generalized) arc-consistent with respect to the csp if and only if for each variable $x_i$,  for each value $v$ in $D(x_i)$, there exist possible values for the other variables of the constraint $c$ that make it true. Thus filtering $c$ consists in removing the values of  $x_1$, $x_2$ \ldots $x_k$ that invalidate that property.
In the previous example, $c_1$ is generalized arc consistent with respect to the given csp. However if we modify the domain of $x_2$ as the singleton $\{1\}$, $c_1$ is not anymore generalized arc consistent  because when $x_1$ has the value 1,  there is no value for $x_2$ that can make the constraint true. In such a case,  a filtering algorithm would remove the value 1 from the domain of $x_1$.


Decomposition of non-binary constraints into equivalent binary constraints
is a subject that has been widely discussed in the CP
community and for quite a long time. A well-known 
transformation for constraint satisfaction problems with finite domains is the Hidden Variable
Encoding (HVE)~\cite{RossiHVE}. recognized as having nice theoretical properties~\cite{MamoulisS01}.
 In HVE, every non-binary constraint is associated with a variable whose domain is
the set of all possible tuples of the original constraint, i.e. the
set of tuples (of values of involved variables in the constraint) that satisfy the constraint. Such a variable
is called a \emph{dual variable} and written $v_c $ if $c$  denotes the
constraint. Thus the variables of the equivalent binary csp are the
variables of the original csp called \emph{original variables}  and the dual variables.
The domains of the original variables remain
identical to their domain in the original csp. Non-binary constraints
do not appear in the binary encoding: they are replaced by \emph{hidden
constraints} between a dual variable and each of the original variables in the constraint
represented by the dual variable.  A hidden constraint enforces the condition that a value of the original variable
must be  the same  as  the  value assigned to it  by the  tuple that is  the value of the  dual
variable~\cite{bacchus02}. In the following we denote them informally as
projections: $proj_1$, $proj_2$, \ldots A mathematical
definition of this transformation (called the \emph{hidden transformation}) can be found in~\cite{bacchus02}
(see Definition 7).

As an illustration, the binary csp resulting from the HVE
transformation  applied on the example presented previously has 10
variables: the 6 original ones and 4 dual variables $v_{c_1}$, $v_{c_2}$,
$v_{c_3}$ and $v_{c_4}$. Domains of original variables remain
identical whereas domains of the dual variables are  such that
\\$D(v_{c_1}) = \{(0,0,1), (0, 1, 0), (1, 0, 0)\}$, \\
$D(v_{c_2}) = \{(0,0,0,1), (0,1,0, 0), (0, 1, 1, 1), (1, 0, 0, 0), (1, 0, 1, 1) , (1, 1, 1, 0)\}$, \\
$D(v_{c_3}) = \{(0,1,0),  (1, 0, 0), (1, 1, 0) , (1, 1, 1)\}$ and\\ $D(v_{c_4}) = \{(0,0,0),  ( 0, 1, 1), (1, 0, 1)\}$.\\ There are 14
binary constraints: the original binary constraint $c_5$ and 13 hidden constraints, 
e.g. $proj_1(v_{c_2}, x_1)$, $proj_3(v_{c_4}, x_6)$. 

\section{Coq Formalisation of HVE Translation}\label{sec:hve}

\subsection{N-ary Constraint Satisfaction Problem}
The Coq formalisation follows the definition given previously, a csp is encoded as a record, of type \coqdocvar{network\_n} (see its definition in the code snippet below) containing a  list of variables, a map from variables to domains (of type \coqdocvar{domain\_n}), represented as lists of values and a list of constraints. Types of variables (\varn) and values  (\valn)  are abstract, they can be further defined either in Coq or in OCaml when extraction is used. 
We expect  \valn{} and \varn{} to be equipped with a strict total order and a decidable equality. 
Constraints (see below the
definition of the type  \constn), either binary or non-binary,
are also abstract but the arity of a constraint is made explicit. It means that the type
of basic constraints (\coqdocvar{basic\_constraint}) is abstract, equipped with a function to get the variables and an abstract interpretation
function (as in \coqbinfd). A non-binary constraint is defined by a value of the abstract type \coqdocvar{OP}, its arity and a list of variables. In order to be as general as possible, we consider extensional constraints as well as intentional ones. In the former case, the semantics is given as a list of acceptable tuples, in the latter case, a  boolean function is expected.
Constraint $c_1$ of the example  given in
Section~\ref{sec:bg} is represented   as 
\coqdocvar{Nary3 p1 [x1 ;  x2 ;  x6]} where \coqdocvar{p1} is associated to the interpretation function 
$f (a, b, c) := a + b + c - 1 = 0$.

\begin{coqdoccode}
\coqdocemptyline
\coqdocnoindent
\coqdockw{Inductive} \coqdocvar{constraint\_n} : \coqdockw{Set} :=\coqdoceol \coqdocnoindent
\ensuremath{|} \coqdocvar{Bin} : \coqdocvar{basic\_constraint} \ensuremath{\rightarrow} \coqdocvar{constraint\_n}\coqdoceol
\coqdocnoindent
\ensuremath{|} \coqdocvar{Nary} :  \coqdocvar{OP} \ensuremath{\rightarrow} \coqdocvar{nat} \ensuremath{\rightarrow} \coqdocvar{list} \coqdocvar{variable\_n}  \ensuremath{\rightarrow} \coqdocvar{constraint\_n}.\coqdoceol
\coqdocemptyline
\coqdocnoindent
\coqdockw{Inductive} \coqdocvar{interpretation} : \coqdockw{Set} :=\coqdoceol
\coqdocnoindent
\ensuremath{|} \coqdocvar{Extension} : \coqdocvar{list} (\coqdocvar{list} \coqdocvar{value\_n}) \ensuremath{\rightarrow} \coqdocvar{interpretation}\coqdoceol\coqdocnoindent
\ensuremath{|} \coqdocvar{Intention} : (\coqdocvar{list} \coqdocvar{value\_n} \ensuremath{\rightarrow} \coqdocvar{bool}) \ensuremath{\rightarrow} \coqdocvar{interpretation}.\coqdoceol
\coqdocemptyline
\coqdocnoindent
\coqdockw{Record} \coqdocvar{network\_n} : \coqdockw{Type} := 
\coqdocindent{1.00em}
\coqdocvar{Make\_cspn} \{\coqdoceol
\coqdocindent{2.00em}
\coqdocvar{CVarsn} : \coqdocvar{list} \coqdocvar{variable\_n} ;\coqdoceol
\coqdocindent{2.00em}
\coqdocvar{Domsn} : \coqdocvar{domain\_n};\coqdoceol
\coqdocindent{2.00em}
\coqdocvar{Cstsn} : \coqdocvar{list} \coqdocvar{constraint\_n}\coqdoceol
\coqdocindent{1.00em}
\}.\coqdoceol
\coqdocemptyline
\end{coqdoccode}

Our formalisation choice requires some extra properties about the input constraints language definition, in particular about the interpretation functions, for example \texttt{basic\_interp} should only be defined for lists of length 2 (should fail for other cases) or the table defining an extensional $k$-ary  constraint should only contain tuples with $k$ components. These requirements can be checked at extraction time, e.g. by testing. They appear in our Coq development as axioms or parameters, in a weak form discovered during the proof of some properties. 
\begin{coqdoccode}
\coqdocemptyline
\coqdocnoindent
\coqdockw{Parameter} \coqdocvar{interp\_op\_length\_extension} : \coqdockw{\ensuremath{\forall}} \coqdocvar{op} \coqdocvar{ar} \coqdocvar{table},\coqdoceol
\coqdocnoindent
\coqdocvar{interp\_op} \coqdocvar{op} \coqdocvar{ar} = \coqdocvar{Extension} \coqdocvar{table} \ensuremath{\rightarrow} 
\coqdockw{\ensuremath{\forall}} \coqdocvar{l}, \coqdocvar{In} \coqdocvar{l} \coqdocvar{table} \ensuremath{\rightarrow} \coqdocvar{length} \coqdocvar{l} = \coqdocvar{ar}.\coqdoceol
\coqdocemptyline
\end{coqdoccode}

The modelling of constraints is as simple as possible. It allows ill-formed constraints. The ability to deal with potentially ill-formed constraints makes the definition of some functions easier. We define well-formedness in a separate way as the predicate named \wfn,  the property is implicitly introduced when needed. 

The predicate \wfn{} specifies the following requirements: 
\begin{itemize}
\item variables in constraints are exactly the ones that are listed in the csp and defined in the domain map;
\item constraints are normalized, meaning that they do not share the same set of variables;
\item basic binary and \nary{} constraints have distinct variables;
\item in the case of a \nary{} constraint of arity $k$ represented by \texttt{Nary op k l}, the length of the list of variables  \coqdocvar{l}  is exactly $k$, with $k$  strictly greater than 2. 
\end{itemize}

An alternative way would have been to use dependent types for constraints, giving to the constructor \nary{} the following  type \texttt{forall n, vector n -> OP n -> constraint\_n} where \texttt{vector n} is the type of lists of length \texttt{n} et \texttt{OP n} the dependent version of the type \texttt{OP}. So a lot of types would become dependent.
Another reason not to use dependent types is that we want to be able to easily define the constraints language in OCaml that does not provide such dependent types. A last reason is that the formalization presented in this paper generalizes the proofs done for ternary constraints~\cite{jfla19} and follows the same line. 

\subsection{Binary Constraint Satisfaction Problem}
A binary csp (as it is encoded in \coqbinfd) has a very similar representation, it is  a record containing a list of variables, a table that maps variables to finite domains and a list of binary constraints.  We define in this subsection the variables, the values and the constraints of a binary csp resulting from the HVE translation.
The type of variables, \coqdocvar{variable}, is defined inductively and reflects that 
variables are either original variables (introduced by the
constructor \coqdocvar{OVar}) or hidden variables (constructor
\coqdocvar{HVar}). The latter variables are defined w.r.t an original constraint. We make explicit this association in the way we build variables. For
example, the hidden variable $v_{c_1}$ of the example given in
Section~\ref{sec:bg} is encoded in Coq as 
\coqdocvar{HVar p1 3  [x1; x2; x6]}.
\begin{coqdoccode}
\coqdocemptyline
\coqdocnoindent
\coqdockw{Inductive} \coqdocvar{variable} := \coqdoceol
\coqdocnoindent
\ensuremath{|} 
\coqdocvar{OVar} : \varn\ \ensuremath{\rightarrow} \coqdocvar{variable}\coqdoceol
\coqdocnoindent
\ensuremath{|} \coqdocvar{HVar} : \coqdocvar{OP} \ensuremath{\rightarrow} \coqdocvar{nat} \ensuremath{\rightarrow} \coqdocvar{list variable\_n} \ensuremath{\rightarrow} \coqdocvar{variable}.\coqdoceol
\coqdocemptyline
\end{coqdoccode}

The type of values, \coqdocvar{value}, is also defined inductively, it
distinguishes raw values, which are the original variables values, from tuples which are the hidden variables values. 

\begin{coqdoccode}
\coqdocemptyline
\coqdocnoindent
\coqdockw{Inductive} \coqdocvar{value} := \coqdoceol
\coqdocnoindent
\ensuremath{|} 
 \coqdocvar{Raw\_value} : \coqdocvar{value\_n} \ensuremath{\rightarrow} \coqdocvar{value}\coqdoceol
\coqdocnoindent
\ensuremath{|} \coqdocvar{Tuple} : \coqdocvar{nat} \ensuremath{\rightarrow} \coqdocvar{tuple} \ \coqdocvar{value}.\coqdoceol
\coqdocemptyline
\end{coqdoccode}

A decidable equality and a strict order are defined for both
types, following from the required equalities and orders on \coqdocvar{value\_n} and
\coqdocvar{variable\_n}.

We can now define the type \coqdocvar{constraint} whose values are the original
binary constraints and the hidden constraints.  In our example, the hidden constraint
between $v_{c_1}$ and the second original variable is represented in
Coq by \coqdocvar{Proj p1 3 [x1; x2; x6] 1 x2}.  We prove the
properties on the constraint language required by \coqbinfd, e.g. any constraint has distinct
variables.

\begin{coqdoccode}
\coqdocemptyline
\coqdocnoindent
\coqdockw{Inductive} \coqdocvar{constraint} : \coqdockw{Set} :=\coqdoceol
\coqdocnoindent
\ensuremath{|} \coqdocvar{Basic} : \coqdocvar{basic\_constraint} \ensuremath{\rightarrow} \coqdocvar{constraint}\coqdoceol
\coqdocnoindent
\ensuremath{|} \coqdocvar{Proj} : \coqdocvar{OP} \ensuremath{\rightarrow} \coqdocvar{nat} \ensuremath{\rightarrow} \coqdocvar{list} \coqdocvar{variable\_n} \ensuremath{\rightarrow} \coqdocvar{nat}  \ensuremath{\rightarrow} \coqdocvar{variable\_n} \ensuremath{\rightarrow} \coqdocvar{constraint}.
\coqdocemptyline
\end{coqdoccode}

\subsection{HVE transformation}\label{subsec:hveCoq}
The Coq function, \coqdocvar{translate\_csp\_n}, that translates a non-binary csp
into a binary csp, closely follows the presentation in
Section~\ref{sec:bg} and the mathematical definition given in~\cite{bacchus02}.  
It uses several intermediate functions, in particular the function \coqdocvar{expand} that
computes the domain of a hidden variable, as a list of tuples, from 
the interpretation function and  the domains of the ordinary variables of 
the constraint corresponding to the hidden variable. The computed domain
contains only the tuples that satisfy the interpretation. In the case of an extensional non-binary constraint, the domain of the corresponding hidden variable is obtained by copying the table given as its interpretation.  
It also uses the function \coqdocvar{cstsnTocsts2} which computes, for a list of
constraints,  the list of original binary and hidden constraints and the list of hidden
variables coupled with their list of tuples computed with the help of \coqdocvar{expand}. 
The ordinary binary constraints
of the original csp and the corresponding domains are just copied
modulo some elementary rewriting. The map containing the domains of the hidden variables is built with
the help of the function \coqdocvar{new\_domain}.

Except some minor differences and the definition of the function \coqdocvar{expand}, the function is similar to the one in the ternary case~\cite{jfla19}.

\begin{coqdoccode}
\coqdocemptyline
\coqdocnoindent
\coqdockw{Definition} \coqdocvar{translate\_csp\_n} \coqdocvar{cspn} :=\coqdoceol
\coqdocnoindent
\coqdockw{match}  (\coqdocvar{cstsnTocsts2} (\coqdocvar{Cstsn} \coqdocvar{cspn}) (\coqdocvar{Domsn} \coqdocvar{cspn}) ) \coqdockw{with}\coqdoceol
\coqdocnoindent
\ensuremath{|} \coqdocvar{None} \ensuremath{\Rightarrow} \coqdocvar{None}\coqdoceol
\coqdocnoindent
\ensuremath{|} \coqdocvar{Some} (\coqdocvar{cs}, \coqdocvar{lvdv}) \ensuremath{\Rightarrow} \coqdocvar{Some} (\coqdocvar{Make\_csp} \coqdoceol
\coqdocindent{1.00em}
(\coqdocvar{List.app} (\coqdocvar{List.map} (\coqdockw{fun} \coqdocvar{x} \ensuremath{\Rightarrow} \coqdocvar{OVar} \coqdocvar{x}) (\coqdocvar{CVarsn} \coqdocvar{cspn})) (\coqdocvar{List.map} \coqdockw{fst} \coqdocvar{lvdv}))\coqdoceol
\coqdocindent{1.00em}
(\coqdocvar{new\_domain} (\coqdocvar{mapn\_to\_raw} (\coqdocvar{Domsn} \coqdocvar{cspn}) (\coqdocvar{CVarsn} \coqdocvar{cspn})) \coqdocvar{lvdv}) \coqdoceol
\coqdocindent{1.00em}
\coqdocvar{cs})\coqdoceol
\coqdocnoindent
\coqdockw{end}.
\coqdoceol
\coqdocemptyline
\end{coqdoccode}

Note that \coqdocvar{translate\_csp\_n} may fail when 
 \coqdocvar{cstsnTocsts2} tries to access the domain of  unknown variables. We prove that if the non-binary csp is
well-formed then the translation does not fail:
%
\begin{coqdoccode}
\coqdocemptyline
\coqdocnoindent
\coqdockw{Lemma} \coqdocvar{network\_inv\_n\_translate\_None\_False} : \coqdockw{\ensuremath{\forall}} \coqdocvar{cspn},\coqdoceol
\coqdocnoindent
\coqdocvar{network\_inv\_n} \coqdocvar{cspn} \ensuremath{\rightarrow}
\ensuremath{\lnot} (\coqdocvar{translate\_csp\_n} \coqdocvar{cspn} =
\coqdocvar{None}).\coqdoceol
\coqdocemptyline
\end{coqdoccode}
We also prove that the binary csp obtained by HVE is well-formed if the original csp is well-formed: 
\begin{coqdoccode}
\coqdocemptyline
\coqdocnoindent
\coqdockw{Lemma} \coqdocvar{translate\_cspn\_network\_inv} : \coqdockw{\ensuremath{\forall}} \coqdocvar{cspn} \coqdocvar{csp},\coqdoceol
\coqdocnoindent
\coqdocvar{network\_inv\_n} \coqdocvar{cspn} \ensuremath{\rightarrow}
\coqdocvar{translate\_cspn} \coqdocvar{cspn} = \coqdocvar{Some}
\coqdocvar{csp} \ensuremath{\rightarrow} \coqdoceol
\coqdocvar{network\_inv} \coqdocvar{csp}.\coqdoceol
\coqdocemptyline
\end{coqdoccode}

\subsection{Focus on tuples and extraction}
Let us focus on the \coqdocvar{expand} function that, in the case of an intentional constraint, computes the set of tuples. It is merely the computation of the cartesian product of $k$ lists if the arity of the constraint is $k$. We first compute the result as a list of lists (of length $k$) representing the tuples. Then we turn these lists into tuples whose type is abstract with the help of an abstract function \coqdocvar{tuple\_from\_list} introduced as a parameter. Yes, this step requires a computational overhead  
but it allows some flexibility at extraction time. For example we can map the type \coqdocvar{tuple} to the OCaml \coqdocvar{array} type in order to benefit from a constant time access. We can also keep  lists by mapping \coqdocvar{tuple\_from\_list} to the identity function.

Besides  \coqdocvar{tuple\_from\_list}, we need 2 other functions: \coqdocvar{tuple\_to\_list} (of type \texttt{nat -> tuple -> value\_n}) and \coqdocvar{proj\_tuple} (of type \coqdocvar{nat -> tuple -> value\_n}). The first one is not used in the translation itself but only in the proofs.  These functions are specified by 3 properties or axioms which are given below:

\begin{coqdoccode}
\coqdocemptyline
\coqdocnoindent
\coqdockw{Parameter} \coqdocvar{tuple\_to\_from\_list} : \coqdockw{\ensuremath{\forall}} \coqdocvar{a} , \coqdoceol
\coqdocnoindent
\coqdocvar{tuple\_to\_list} (\coqdocvar{length} \coqdocvar{a}) (\coqdocvar{tuple\_from\_list} \coqdocvar{a}) = \coqdocvar{a}.\coqdoceol
\coqdocemptyline
\coqdocnoindent
\coqdockw{Parameter} \coqdocvar{proj\_tuple\_nth\_error} : \coqdockw{\ensuremath{\forall}} \coqdocvar{n} \coqdocvar{n0} \coqdocvar{t} \coqdocvar{v0},\coqdoceol
\coqdocnoindent
\textcolor{blue}{\coqdocvar{n} > 0} \ensuremath{\rightarrow}
\coqdocvar{n0} <  \coqdocvar{n} \ensuremath{\rightarrow}\coqdoceol
\coqdocnoindent
\coqdocvar{proj\_tuple} \coqdocvar{n0} \coqdocvar{t} = \coqdocvar{v0} \ensuremath{\leftrightarrow} \coqdocvar{nth\_error} (\coqdocvar{tuple\_to\_list} \coqdocvar{n} \coqdocvar{t}) \coqdocvar{n0} = \coqdocvar{Some} \coqdocvar{v0}.\coqdoceol
\coqdocemptyline
\coqdocnoindent
\coqdockw{Parameter} \coqdocvar{length\_tuple\_to\_list} : \coqdockw{\ensuremath{\forall}} \coqdocvar{n} \coqdocvar{t}, \coqdoceol
\coqdocnoindent
\coqdocvar{length} (\coqdocvar{tuple\_to\_list} \coqdocvar{n} \coqdocvar{t}) = \coqdocvar{n}.\coqdoceol
\coqdocemptyline
\end{coqdoccode}

In order to gain some more confidence when we extract OCaml code from the Coq code, we have tested these 3 properties using the QuickChick property testing tool for Coq programs~\cite{Denes14qc} with  10000 test cases randomly generated. It allowed the discovery of a missing hypothesis (the blue one in the second statement).

An alternative could be to use primitive persistent arrays in the Coq code for implementing tuples (without going through intermediate lists). The type of such arrays is axiomatized (in the \texttt{PArray} module).  Primitive arrays  internally are implemented using a persistent data structure. This has been very recently integrated into the current version of Coq. It was previously available as a separate implementation~\cite{ArmandGST10}.  We plan to experiment with these primitive arrays. The cartesian product of domains implemented in the \coqdocvar{expand} function is a bit more complicated when dealing with arrays. 

A last proposition could be to implement tuples as finite functions, and then to  use the coq library proposed by Sakaguchi
 in~\cite{Sakaguchi20} to extract these tuples to OCaml arrays. 

\subsection{Correctness of the HVE translation}
To prove the correctness of the translation, we prove that satisfiability is preserved by the HVE translation. Two related properties are illustrated below.

A solution is defined as usual as an assignment of the csp variables
which is total, valid (i.e. values are compatible with the  domains) and locally consistent (i.e. making each constraint satisfied). It is implemented as a
map from variables to values. A solution of a non-binary csp (resp. a binary encoding csp) is characterized by the predicate \coqdocvar{solution\_n} (resp. \coqdocvar{solution}). 

Lemma \coqdocvar{translate\_nosol} states that if the original non-binary csp is
 UNSAT (i.e. it admits no solution) then the binary encoding is also
 UNSAT.  Lemma \coqdocvar{translate\_complete} explains that if the non-binary
 csp admits a solution, \coqdocvar{a\_n}, then its mapping to the hidden and original
 variables (computed by the function \coqdocvar{translate\_sol\_n}) is a solution of the binary
 encoding. 
\begin{coqdoccode}
\coqdocemptyline 
\coqdocnoindent
\coqdockw{Lemma} \coqdocvar{translate\_nosol}: \coqdockw{\ensuremath{\forall}} \coqdocvar{cspn} 
\coqdocvar{csp}
,\coqdoceol
\coqdocindent{1.00em}
\coqdocvar{network\_inv\_n} \coqdocvar{cspn}  \ensuremath{\rightarrow} 
\coqdocvar{translate\_csp\_n} \coqdocvar{cspn} = \coqdocvar{Some} \coqdocvar{csp} \ensuremath{\rightarrow} \coqdoceol
\coqdocindent{1.00em}
(\coqdockw{\ensuremath{\forall}} \coqdocvar{a},
\ensuremath{\lnot} (\coqdocvar{solution} \coqdocvar{a} \coqdocvar{csp}))
\ensuremath{\rightarrow} \coqdockw{\ensuremath{\forall}} \coqdocvar{an}, \ensuremath{\lnot} (\coqdocvar{solution\_n}  \coqdocvar{an} \coqdocvar{cspn}).\coqdoceol
\coqdocemptyline
\coqdocnoindent
\coqdockw{Lemma} \coqdocvar{translate\_complete}:  \coqdockw{\ensuremath{\forall}} \coqdocvar{an} \coqdocvar{cspn} \coqdocvar{csp},\coqdoceol
\coqdocindent{1.00em}
\coqdocvar{network\_inv\_n} \coqdocvar{cspn} \ensuremath{\rightarrow}
\coqdocvar{translate\_csp\_n} \coqdocvar{cspn} = \coqdocvar{Some} \coqdocvar{csp} \ensuremath{\rightarrow}\coqdoceol
\coqdocindent{1.00em}
\coqdocvar{solution\_n} \coqdocvar{an} \coqdocvar{cspn} \ensuremath{\rightarrow} \coqdocvar{solution} (\coqdocvar{translate\_sol\_n} \coqdocvar{an} \coqdocvar{cspn}) \coqdocvar{csp}.\coqdoceol
\coqdocemptyline
\end{coqdoccode}

\section{Brief Presentation of the Formally Verified Solver \coqbinfd} \label{sec:coqbinfd}
In this section we briefly describe the binary solver \coqbinfd{} 
that we want to reuse. For more details please consult~\cite{CDG12}.
An important point in this development and crucial for the present
work is its genericity. In the following we mainly emphasize the requirements
upon the generic parameters. The solver is indeed parameterized by the type
of variables (\coqdocvar{variable}) and values (\coqdocvar{value}) and also by the
constraint language (\coqdocvar{constraint}). In Coq,
these types are abstract, assumed to accept a decidable equality. 
It is also assumed that the semantics of the constraints is given by
an interpretation function as a Boolean function
of the values of its two variables and a function 
 that retrieves, for any constraint,  its two
variables. 
So a constraint is abstracted as a relation over two
distinct variables, represented by an interpretation predicate.
These types and functions must be defined either in Coq or
directly in OCaml in order to use the extracted solver in a particular
context. Here  they will be given Coq  concrete values 
according to HVE.

A csp  is defined  as a record of type
\coqdocvar{network\_csp} consisting of a
finite list of variables (\coqdocvar{CVars} field), a finite list of constraints (\coqdocvar{Csts})
and a map (\coqdocvar{Doms}) associating each domain with its variable, here a
finite  list of values. A predicate (\texttt{network\_inv}) specifies the
well-formedness of a csp: the entries of the domain map are exactly
the variables of the csp, 
variables appearing in the constraints are exactly those declared, 
 constraints are normalized and finally the two variables of any
 constraint are distinct. 

The solving process is based on arc-consistency, it implements a generic version of the propagation
algorithm AC3~\cite{AI77}, allowing the use of  AC2001~\cite{AI77}.
However in our work, it is transparent, the binary solver
being used as a black-box. 

\section{Extension of the Solver \coqbinfd{} to Non-binary Constraints}\label{sec:extension}

\begin{figure}
\centering
\begin{tikzpicture}[scale=0.38]
\node (orig)  at (0,0) {};
\node [rectangle,fill=blue!20,draw,inner sep=0.3cm,rounded corners=0.3cm] (csp)  at (4,0) {cspn};
\node [rectangle, inner color=white, outer color=red!15 , inner sep=0.3cm] (hne) at ( 9.5, 0) {HVE};
\node [rectangle , fill=blue!20,draw,inner sep=0.3cm,rounded corners=0.3cm] (cspbin) at (15, 0) {csp2};
\node [rectangle, inner color=white, outer color=red!15 , inner sep=0.3cm] (solverbin) at (22, 0) {\coqbinfd};
\node [rectangle , fill=blue!20,draw,inner sep=0.3cm,rounded corners=0.3cm] (solbin) at (22, -4) {sol2 / unsat};
\node [rectangle, inner color=white, outer color=red!15 , inner sep=0.3cm] (translate) at (15, -4) {translate};
\node [rectangle , fill=blue!20,draw,inner sep=0.3cm,rounded corners=0.3cm] (sol) at (8, -4) {soln / unsat};

\draw[->] (csp) -- (hne);
\draw[->] (hne) -- (cspbin);
\draw[->] (cspbin) -- (solverbin);
\draw[->] (solverbin) -- (solbin);
\draw[->] (solbin) -- (translate);
\draw[->] (translate) -- (sol);
\end{tikzpicture} 
\caption{Behavior of the Non-binary Solver} \label{fig:combination}
\end{figure}
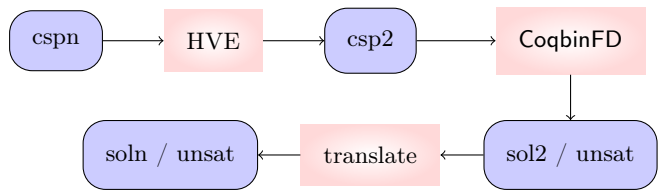

We propose to build a constraint solver able to deal with binary and non-binary constraints by extending the \coqbinfd{} solver (whose main function is the \coqdocvar{solve\_csp} function) with the HVE translation acting as a pre-processor and the solution translation  acting as a post-processing. The different steps are illustrated on Fig.~\ref{fig:combination}. 

The extended solver is mainly embodied by the following \coqdocvar{solve\_n} function which follows the steps of Fig.~\ref{fig:combination} and is built using the tactic \texttt{Program}~\cite{Sozeau06} (as its counterpart in \coqbinfd):
\begin{coqdoccode}
\coqdocemptyline
\coqdocnoindent
\coqdockw{Program Definition} \coqdocvar{solve\_n} 
(\coqdocvar{cspn} : \coqdocvar{network\_n}) 
(\coqdocvar{Hn}: \coqdocvar{network\_inv\_n} \coqdocvar{cspn}) \coqdoceol 
: \{\coqdocvar{res} : \coqdocvar{option} (\coqdocvar{assign\_n} ) | \coqdocvar{result\_n\_ok} \coqdocvar{res} \coqdocvar{csp}\} :=\coqdoceol
\coqdocnoindent
\coqdockw{match} (\coqdocvar{translate\_csp\_n} \coqdocvar{cspn}) \coqdockw{with}\coqdoceol
\coqdocindent{1.00em}
\coqdocvar{None} \ensuremath{\Rightarrow} \coqdocvar{None} \coqdoceol
\coqdocnoindent
\ensuremath{|} \coqdocvar{Some} \coqdocvar{csp} \ensuremath{\Rightarrow} 
\coqdockw{match} (\coqdocvar{solve\_csp} \coqdocvar{csp} \coqdocvar{\_}) \coqdockw{with}\coqdoceol
\coqdocindent{6.90em}
\coqdocvar{None} \ensuremath{\Rightarrow} \coqdocvar{None} \coqdoceol
\coqdocindent{6.50em}
\ensuremath{|} \coqdocvar{Some} \coqdocvar{a} \ensuremath{\Rightarrow} \coqdocvar{Some} (\coqdocvar{translate\_sol} \coqdocvar{a} \coqdocvar{cspn.CVarsn})\coqdoceol
\coqdocindent{6.50em}
\coqdockw{end}\coqdoceol
\coqdocnoindent
\coqdockw{end}.\coqdoceol
\coqdocemptyline
\end{coqdoccode}

The type of the result is a kind of subtype \textit{à la PVS}, it describes not only the type of the computed result \coqdocvar{res} (\coqdocvar{None} or  a Nary solution) but it also contains a proof that the result is sound (specified by the predicate \coqdocvar{result\_n\_ok}), i.e. if the result is \coqdocvar{None} then the original csp has no solution and if it is \coqdocvar{Some a}, then \coqdocvar{a} is a solution of the original csp.  
This definition generates proof obligations that correspond to the expected properties of the result. Another proof obligation comes from the underscore appearing in the call of \coqdocvar{solve\_csp} that expects as a third argument a proof that its second argument is well-formed. This proof obligation is solved by the  lemma \coqdocvar{translate\_cspn\_network\_inv} shown previously in Subsection~\ref{subsec:hveCoq}. 

Completeness of the extended solver is also proved. It follows from the completeness of \coqbinfd{} and from the properties of 
the \coqdocvar{translate\_sol\_n} function regarding solutions. 

The main task to extend \coqbinfd{} to Nary constraints  is to provide the binary encoding exactly as it is expected by \coqbinfd{}. As this solver is generic in the input constraint language, the task was made easier.

After extraction, we ran  the extended solver (with the AC3 instance of \coqbinfd) to solve some problems. First we have used
it with binary and ternary csps, for non-regression testing. The time overhead is
not significant.
We also solved some problems, intensional and extensional ones, from the XCSP2.1 library~\cite{xcsp} where csps are represented as XML definitions. For example the problem named \texttt{normalized\_g\_4x4} with 16 variables with $\{0,1\}$ as domain and 15 constraints with arity from 3 to 5 is solved in 0.0033 sec on a laptop
  (2,3 GHz Intel Core i5 8 Go 2133 MHz LPDDR3) whereas for the problem known as \texttt{normalized-graceful-K2-P3} with 15 variables (whose  domain is either 0..9 or 1..9) and 60 constraints,  9 of them being  ternary and the rest being binary, we obtain a solution in 2.8 sec.
  
The manual transcription of XCSP2.1 problems in OCaml is however tedious and error prone. Our solver could be completed with a tool allowing the translation of  XML definitions into OCaml or Coq definitions.  Following Stergiou and Samaras in~\cite{StergiouS05}, we could obtain a better efficiency by using specialized arc consistency and search algorithms for the binary encodings requiring to further prove some variants for propagation and exploration algorithms. 



\section{Conclusion}
In this paper we have formalized in Coq the well-known Hidden Variable Encoding that performs the translation of a non-binary constraint satisfaction problem into an equivalent binary  constraint satisfaction problem. This translation is used to extend the \coqbinfd{} solver, developed in Coq several years ago. The Coq code is available at \url{www.ensiie.fr/~dubois/HVE_nary}. From the whole Coq development, an OCaml executable solver can be extracted. It can be considered as a reference implementation and used to test other solvers, for example the FaCiLe OCaml constraint library~\cite{facile01}. 

\bibliographystyle{abbrv}
\bibliography{cp18}

\end{document}